\documentstyle[aps,amssymb]{revtex}

\begin{document}
\author{R. A. Serota}
\address{Department of Physics\\
Mail Location \# 0011\\
University of Cincinnati\\
Cincinnati, OH\ 45221-0011\\
serota@physics.uc.edu}
\title{VanVleck Response Of A Two-Level System And\\
Mesoscopic Orbital Magnetism Of Small Metals}
\date{first draft 11/17/99}
\maketitle

\begin{abstract}
We evaluate the mean value of the van Vleck response of a two-level system
with level spacing distribution and argue that it describes the orbital
magnetism of small conducting particles.
\end{abstract}

\section{Introduction}

The significance of the few-level physics for small metal particles has long
been realized\cite{GE},\cite{DMS}. At temperatures smaller than the mean
level spacing at the Fermi level, $T\lesssim \Delta $, the spin
susceptibility, for instance, is determined by the (spin-flip) transitions
to the first unoccupied state for the particles that contain even number of
electrons and is typically much smaller than the Pauli susceptibility. For
the odd-electron particles, it is given by the Curie susceptibility. This is
in contrast to the $T\gg \Delta $ (and the bulk) case where the leading term
is given by the Pauli susceptibility\cite{SS1}. Moreover, due to the
exponential activation factors associated with transitions to unoccupied
states, one should expect extremely broad distribution of susceptibility
values and, therefore, large variations from particle to particle\cite{SS2}.

The orbital magnetic response of small metal particles, on the other hand,
does not easily reduce to a few-level problem. While it does turn out to be
essentially a Fermi level property for $T\gg \Delta $\cite{SS3} and in the
bulk systems\cite{LL1}-\cite{F}, a priori it appears as a property of the
entire Fermi sea since all the levels are perturbed by the magnetic field.
The two competing contributions to orbital magnetism are the precession\
diamagnetism and the polarization (van Vleck) paramagnetism\cite{V}. While
no rigorous description of the orbital magnetism exists for $T\lesssim
\Delta $, it is nonetheless expected to be a Fermi level property as well
due to the large Fermi sea cancellations between the diamagnetic and
paramagnetic contributions. However, at the Fermi level the van Vleck part
of the orbital response should be dominant and also the one that is very
sensitive to variations of the energy level spacings. Consequently, we study
a model wherein it is assumed that among all the ''virtual transitions''
between the ground and the excited states of the Fermi sea that are
responsible for the van Vleck response, the one that determines the particle
susceptibility is a single-electron ''transition'' from the last occupied to
the first unoccupied state.

Imry has conjectured\cite{I} that, since the electrons in a metal particle
should be considered as a canonical ensemble\cite{K}, one can describe the
particle response in terms of the difference between the magnetic energies
of the canonical and grand canonical ensemble $F_{H}-\Omega _{H}$, the
latter being the case for a bulk system and relatively well understood. In
the limiting case of $T\lesssim \Delta $, one should expect the difference
to be due to a single electron level crossing the position of the chemical
potential in the equivalent grand canonical ensemble. However, the formalism
based on this approach works only in the limit of $T\gg \Delta $, as
explained in Refs.\cite{SS1},\cite{SS3}, in which case $F-\Omega =\Delta
\delta N^{2}/2$ where the particle number fluctuation in the equivalent
grand canonical ensemble is actually much larger than one, $\delta N^{2}\sim
T/\Delta \gg 1$\cite{LL1}. In this limit, the perturbation theory can be used%
\cite{AS}. An improvement on this approach\cite{SS3} involves using the
exact, non-perturbative level correlation function and, in fact, yields a
saturation value of the magnetic energy as $T$ approaches $\Delta $ from
above.

In what follows, we first give an argument in support of the large Fermi sea
cancellation between the diamagnetic and paramagnetic contributions to the
magnetic response. We then evaluate the two-level van Vleck response for the
Fermi level and first unoccupied level. This involves averaging of the
inverse energy spacing using the Gaussian Orthogonal Ensemble (GOE)
statistics for energy eigenvalues and calculation of the magnetic dipole
moment. The latter can be estimated using a semiclassical argument but also
rigorously evaluated by considering the magnetic dipole absorption. We
compare our result with the expression for the orbital magnetic response
obtained in the limit $T>\Delta $. We restrict our analysis to 2D diffusive
particles (disks, narrow strips, and rings) and use the units where $c=\hbar
=1$. We also neglect any sample-specific effects (fluctuations), except when
specifically mentioned, and all the quantities in consideration are presumed
disorder-averaged.

\section{Fermi Sea Cancellation Of Van Vleck Paramagnetism And Precession
Diamagnetism}

The van-Vleck energy of a two-level system in the magnetic field ${\bf H}%
=H_{0}\widehat{{\bf z}}$ is 
\begin{equation}
\delta \epsilon _{vV}^{\left( 1\right) }=\frac{\left| \left\langle i\right| 
\widehat{M}_{z}\left| f\right\rangle \right| ^{2}H_{0}^{2}}{\varepsilon
_{i}-\varepsilon _{f}}\equiv \frac{\left| \widehat{M}_{if}\right|
^{2}H_{0}^{2}}{\varepsilon _{i}-\varepsilon _{f}}  \label{E_vV-1}
\end{equation}
and the total van Vleck energy is given by\cite{V} 
\begin{equation}
\epsilon _{vV}^{\left( tot\right) }=\frac{1}{2}\sum_{i,f}\left| \widehat{M}%
_{if}\right| ^{2}H_{0}^{2}\frac{n_{i}-n_{f}}{\varepsilon _{i}-\varepsilon
_{f}}  \label{E_vV-tot}
\end{equation}
where $n_{i}=\theta \left( -\varepsilon _{i}\right) $. In the case of a
continuous spectrum, we can rewrite eq. (\ref{E_vV-tot}), using $%
dn_{i}/d\varepsilon _{i}=-\delta \left( \varepsilon _{i}\right) $, as 
\begin{equation}
\epsilon _{vV}^{\left( tot\right) }=-\frac{1}{2}\upsilon
H_{0}^{2}\sum_{f}\left| \widehat{M}_{0f}\right| ^{2}=-\frac{1}{2}\upsilon
\left\langle 0\right| M_{z}^{2}\left| 0\right\rangle H_{0}^{2}=-\frac{%
\upsilon e^{2}v_{F}^{2}\left\langle 0\right| r_{\perp }^{2}\left|
0\right\rangle H_{0}^{2}}{16}  \label{E_vV_final-tot}
\end{equation}
where $\upsilon \equiv \left\langle \upsilon \left( 0\right) \right\rangle $
is the mean level density at the Fermi level, $\upsilon \left( \varepsilon
\right) $ is the level density at energy $\varepsilon $. We have also used
the fact that the magnetic moment at the Fermi level can be evaluated
semiclassically, that is 
\begin{equation}
{\bf M}=\frac{e}{2}{\bf r\times v}_{F}  \label{M}
\end{equation}
where ${\bf r}$ is the classical position vector of an electron and ${\bf v}%
_{F}$ is the Fermi velocity. In evaluating the coefficient in eq. (\ref
{E_vV_final-tot}), we averaged over the angle between ${\bf r}$ and ${\bf v}%
_{F}$.

The single-level diamagnetic energy diamagnetic energy is given by 
\begin{equation}
\delta \epsilon _{diam}^{\left( 1\right) }=\frac{e^{2}\left\langle i\right|
r_{\perp }^{2}\left| i\right\rangle H_{0}^{2}}{8m}  \label{E_diam-1}
\end{equation}
and the total diamagnetic energy of the Fermi sea is given by\cite{V} 
\begin{equation}
\delta \epsilon _{diam}^{\left( tot\right) }=\sum_{i}\frac{e^{2}\left\langle
i\right| r_{\perp }^{2}\left| i\right\rangle H_{0}^{2}}{8m}n_{i}
\label{E_diam-tot}
\end{equation}
It is reasonable to conjecture that the disorder-averaged value $%
\left\langle i\right| r_{\perp }^{2}\left| i\right\rangle $ is $i$
-independent which yields, upon converting to integration for a continuous
spectrum, 
\begin{equation}
\delta \epsilon _{diam}^{\left( tot\right) }=\frac{\upsilon
e^{2}v_{F}^{2}\left\langle 0\right| r_{\perp }^{2}\left| 0\right\rangle
H_{0}^{2}}{16}  \label{E_diam_final-tot}
\end{equation}
and is the same as eq. (\ref{E_vV_final-tot}), with the opposite sign. This
confirms the Fermi sea cancellation between the van Vleck paramagnetism and
precession diamagnetism\footnote{%
This argument can be carried over, with minor modifications, to 3D.}$^{,}$%
\footnote{%
For a disk of radius $R$, $\left\langle 0\right| r_{\perp }^{2}\left|
0\right\rangle $ can be evaluated as the area average and equals to $R^{2}/2$%
.}. However, this derivation does not account for the quantum effects beyond
the existence of the Fermi sea. Consequently, it is understood that the
cancellation is not exact and that a Fermi level contribution is not
accounted for in the present approximation. The nature of this contribution
should depend on whether the chemical potential or the number of particles
is fixed. In the former case, one expects a Landau response, as in bulk
systems, while in the latter we make an ansatz of a two-level van Vleck
response which involves the last occupied (Fermi) level and the first
unoccupied level.

The situation is more complex in a strictly discrete level case where we
will give only an order-of-magnitude argument. The first principles estimate
of $\left| \widehat{M}_{if}\right| ^{2}$ in the diffusive regime is based on
the idea first proposed by Shapoval\cite{S} and, later, applied by Gor'kov
and Eliashberg\cite{GE} to small metal particles. Namely, we use the
semiclassical approach to write 
\begin{equation}
\left| \widehat{M}_{if}\right| ^{2}=\frac{1}{\pi \upsilon }\overline{%
\int_{0}^{\infty }d\tau \exp \left[ \left( \varepsilon _{i}-\varepsilon
_{f}\right) \tau \right] M\left( t\right) M\left( t+\tau \right) }
\label{M_if_sc}
\end{equation}
where the bar denotes averaging over all classical trajectories. The
correlation time scale $\tau $ for $M\left( t\right) $ is given by 
\begin{equation}
\tau \sim \frac{\ell ^{2}}{D}  \label{tau}
\end{equation}
where $\ell =v_{F}\tau $ is the electron mean-free-path and $D=v_{F}^{2}$ $%
\tau /2$ is the diffusion coefficient. This is because the directions of $%
{\bf v}_{F}$ are uncorrelated after such time. Also, for $\varepsilon
_{f}-\varepsilon _{i}>\tau ^{-1}$ the exponential term becomes oscillatory.
The scale of ${\bf r}$ is the relevant sample dimension $a$ and we find, 
\begin{equation}
\left| \widehat{M}_{if}\right| ^{2}\approx \frac{1}{\pi \upsilon }\overline{%
\int_{0}^{\infty }d\tau M\left( t\right) M\left( t+\tau \right) }\sim \frac{%
e^{2}\ell ^{2}v_{F}^{2}a^{2}}{\upsilon D}\sim \frac{e^{2}Da^{2}}{\upsilon }%
\sim \mu _{B}^{2}\left( \varepsilon _{F}\tau \right)  \label{M_if-est}
\end{equation}
where $\mu _{B}$ is the Bohr magneton and $\varepsilon _{F}$ is the Fermi
energy. Consequently, the order of magnitude value of the two-level van
Vleck response is obtained from eq. (\ref{E_vV-1}) as 
\begin{equation}
\delta \epsilon _{vV}^{\left( 1\right) }\sim -\frac{\left| \widehat{M}%
_{if}\right| ^{2}H_{0}^{2}}{\Delta }\sim -\frac{\mu _{B}^{2}H_{0}^{2}}{%
\Delta }\left( \varepsilon _{F}\tau \right) \sim -\left| \chi _{L}\right|
H_{0}^{2}A\left( \varepsilon _{F}\tau \right)  \label{E_vV_1-est}
\end{equation}
where $\chi _{L}$ is the Landau susceptibility\cite{LL1} and $A\sim a^{2}$
is the sample area. The total van Vleck energy can be estimated by
multiplying $\delta \epsilon _{vV}^{\left( 1\right) }$ by $\left( \tau
\Delta \right) ^{-1}$, which determines the range of the integral in eq. (%
\ref{M_if_sc}), and we find 
\begin{equation}
\epsilon _{vV}^{\left( tot\right) }\sim -\frac{\mu _{B}^{2}H_{0}^{2}}{\Delta 
}\frac{\varepsilon _{F}}{\Delta }\sim -\left| \chi _{L}\right| H_{0}^{2}A%
\frac{\varepsilon _{F}}{\Delta }  \label{E_vV_est-tot}
\end{equation}
Since $\left\langle 0\right| r_{\perp }^{2}\left| 0\right\rangle \sim a^{2}$
, this is in qualitative agreement with eq. (\ref{E_vV_final-tot}). Notice
also that eq. (\ref{E_diam-1}) yields 
\begin{equation}
\delta \epsilon _{diam}^{\left( 1\right) }\sim \frac{\mu _{B}^{2}H_{0}^{2}}{%
\Delta }\sim \left| \chi _{L}\right| H_{0}^{2}A  \label{E_diam_est-1}
\end{equation}
for a single-level contribution to the precession diamagnetism.

\section{Van Vleck Paramagnetism In The Two-Level Model}

Turning again to eq. (\ref{E_vV-1}), where now $\varepsilon _{i}$ is the
energy of the last occupied state (Fermi level) and $\varepsilon _{f}$ is
the energy of the first unoccupied state, we find, averaging with the
Wigner-Dyson distribution\cite{BFFMPW}, 
\begin{equation}
\delta \epsilon =-s\left| \widehat{M}_{if}\right| ^{2}H_{0}^{2}\frac{\pi }{%
2\Delta }\int_{0}^{\infty }\exp \left( -\frac{\pi x^{2}}{4}\right) dx=-\frac{%
\pi \upsilon }{2}\left| \widehat{M}_{if}\right| ^{2}H_{0}^{2}
\label{E_vV-GOE}
\end{equation}
Here $s$ is the level degeneracy ($s=2$, on the account of spin) and $%
\upsilon =s\Delta ^{-1}$. We have already estimated the matrix element using
the semiclassical argument. However, a precise derivation of $\left| 
\widehat{M}_{if}\right| ^{2}$ in the diffusive regime can be done by means
of evaluation of the low-frequency magneto-dipole absorption in the field $%
{\bf H}=H_{0}\exp \left( -i\omega t\right) \widehat{{\bf z}}$. In the case
of the continuous energy spectrum (for instance, when the level broadening $%
\gamma $ is larger than $\Delta $ ), the quantum-mechanical expression for
the absorption should yield, up to small corrections, the classical value%
\cite{WMW}. This is in complete analogy with the electric-dipole absorption
(barring screening effects for the latter), which is discussed in detail in
Ref.\cite{SG}. Since $\left| \widehat{M}_{if}\right| ^{2}$ enters into the
quantum-mechanical expression (see below), it can be extracted by equating
with the classical value of the absorption.

The classical absorption is readily evaluated according to\cite{LL2}

\begin{equation}
Q_{class}=\frac{1}{2}\omega 
\mathop{\rm Im}%
\left\{ M_{z}^{*}H_{0}\right\}  \label{Q_class}
\end{equation}
where 
\begin{eqnarray}
M_{z}^{\left( disk\right) } &=&-\frac{AH_{0}}{4\pi }\left( 1-\frac{2}{%
\varkappa R}\frac{J_{1}\left( \varkappa R\right) }{J_{0}\left( \varkappa
R\right) }\right) \approx -\frac{AH_{0}}{32\pi }\left( \varkappa R\right)
^{2}\text{, }A=\pi R^{2}  \label{M_class_disk} \\
M_{z}^{\left( strip\right) } &=&-\frac{AH_{0}}{8\pi }\left( 1-\frac{\tan
\left( \varkappa L_{x}/2\right) }{\varkappa L_{x}/2}\right) \approx -\frac{
AH_{0}}{96\pi }\left( \varkappa L_{x}\right) ^{2}\text{, }A=L_{x}L_{y}
\label{M_class_strip}
\end{eqnarray}
for a disk of radius $R$ and for a narrow metal strip, such that $L_{x}\ll
L_{y}$, respectively. Here 
\begin{equation}
\varkappa =\frac{1+i}{\delta }\text{, }\delta =\frac{1}{\sqrt{2\pi \sigma
\omega }}  \label{skin}
\end{equation}
and $\sigma $ is the Boltzmann conductivity. It is assumed that the
frequency is such that $\delta \gg R,L_{x}$. Combining eqs. (\ref{Q_class})-(%
\ref{skin}), we obtain 
\begin{eqnarray}
Q_{class}^{\left( disk\right) } &=&\frac{\omega ^{2}H_{0}^{2}R^{2}A\sigma }{
16}  \label{Q_class_disk-final} \\
Q_{class}^{\left( strip\right) } &=&\frac{\omega
^{2}H_{0}^{2}L_{x}^{2}A\sigma }{48}  \label{Q_class_strip-final}
\end{eqnarray}
for the absorption, respectively, in the disk and the strip.

The quantum absorption, for the continuous spectrum, can be evaluated (in
complete analogy with the electric-dipole absorption\cite{SG}) by means of
the Fermi golden rule and we find 
\begin{equation}
Q_{cont}=\frac{\pi }{2}\omega ^{2}\upsilon ^{2}H_{0}^{2}\left| \widehat{M}
_{if}\right| ^{2}\frac{\left\langle \upsilon \left( 0\right) \upsilon \left(
\omega \right) \right\rangle }{\upsilon ^{2}}\approx \frac{\pi }{2}\omega
^{2}\upsilon ^{2}H_{0}^{2}\left| \widehat{M}_{if}\right| ^{2}
\label{Q_quant-final}
\end{equation}
where the small quantum corrections of order $\Delta ^{2}/\gamma ^{2}$ (or $%
\Delta ^{2}/\omega ^{2}$ if $\omega >\gamma $)\cite{AS} are neglected. For
the electric-dipole absorption\cite{GE},\cite{SG}, the quantity
corresponding to $\widehat{M}_{if}$ is the electric dipole matrix element $%
\widehat{P}_{if}$ . The latter can be evaluated from first principles in the
diffusive approximation and the classical and quantum results can be
evaluated independently and are equal in the considered limit. On physical
grounds, we can assume that this is also true for the magnetic-dipole
absorption. Consequently, we equate the r.h.s. of eqs. (\ref
{Q_class_disk-final}), (\ref{Q_class_strip-final}) with that of eq. (\ref
{Q_quant-final}) and, with the use of $\sigma =e^{2}\upsilon D/A$, we obtain 
\begin{eqnarray}
\frac{\pi }{2}\upsilon \left| \widehat{M}_{if}^{\left( disk\right) }\right|
^{2} &=&\frac{e^{2}DR^{2}}{16}  \label{M_if_disk} \\
\frac{\pi }{2}\upsilon \left| \widehat{M}_{if}^{\left( strip\right) }\right|
^{2} &=&\frac{e^{2}DL_{x}^{2}}{48}  \label{M_if_strip}
\end{eqnarray}
for the disk and the strip respectively.

Substituting thus found value of $\left| \widehat{M}_{if}\right| ^{2}$ in
eq. (\ref{E_vV-GOE}), we find (for $s=2$) the following result for the van
Vleck energy: 
\begin{eqnarray}
\delta \epsilon ^{\left( disk\right) } &=&-\frac{1}{16}e^{2}DR^{2}H_{0}^{2}
\label{E_vV_disk-final} \\
\delta \epsilon ^{\left( strip\right) } &=&-\frac{1}{48}%
e^{2}DL_{x}^{2}H_{0}^{2}  \label{E_vV_strip-final}
\end{eqnarray}
for the disk and the strip respectively. Eqs. (\ref{E_vV_disk-final}) and (%
\ref{E_vV_strip-final}) should be compared with the magnetic part of the
energy obtained for $T>\Delta $ in the so called ''mixed approximation''\cite
{SS3} wherein the exact, non-perturbative level correlation function is used
yet the large particle-number fluctuation in the equivalent grand canonical
ensemble is assumed also, the latter being true only for $T\gg \Delta $, 
\begin{equation}
\delta \epsilon _{>}=-\frac{1}{2\pi }{{\tau }_{H}^{-1}}  \label{E_greater}
\end{equation}
The procedure for the evaluation of ${{\tau }_{H}^{-1}}$ using the gauge
where the vector potential is tangential to the surface is described in Ref.%
\cite{OZS} and we find 
\begin{eqnarray}
\delta \epsilon _{>}^{\left( disk\right) } &=&-\frac{1}{4\pi }%
e^{2}DR^{2}H_{0}^{2}  \label{E_greater_disk} \\
\delta \epsilon _{>}^{\left( strip\right) } &=&-\frac{1}{6\pi }%
e^{2}DL_{x}^{2}H_{0}^{2}  \label{E_greater_strip}
\end{eqnarray}
for the disk and the strip respectively.

We also mention the Aharonov-Bohm response of a narrow ring threaded by the
flux $\phi $, where the derivation is particularly simple. Namely, the van
Vleck energy in this case is given by 
\begin{equation}
\delta \epsilon _{if}=-\frac{\left| \widehat{v}_{if}\right| ^{2}}{%
\varepsilon _{i}-\varepsilon _{f}}\left( \frac{e\phi }{2\pi R}\right) ^{2}
\label{E_vV_ring-1}
\end{equation}
where $\left| \widehat{v}_{if}\right| ^{2}$ is evaluated by considering
absorption in the alternating flux $\phi \exp \left( -i\omega t\right) $
which generates the electric field according to the Lenz's law that, in
turn, produces the Boltzmann current density proportional to $v$.
Consequently, we find 
\begin{equation}
\left| \widehat{v}_{if}\right| ^{2}=\frac{D}{4\pi \upsilon }
\label{v_if_ring}
\end{equation}
and, averaging over the level spacing, 
\begin{equation}
\delta \epsilon ^{\left( ring\right) }\sim -\frac{D}{8}\left( \frac{e\phi }{%
\pi R}\right) ^{2}  \label{E_vV_ring-final}
\end{equation}
The latter has the same parametric dependence as 
\begin{equation}
\delta \epsilon _{>}^{\left( ring\right) }=-\frac{1}{2\pi }{{\tau }_{H}^{-1}=%
}-\frac{D}{2\pi }\left( \frac{e\phi }{\pi R}\right) ^{2}
\label{E_greater_ring}
\end{equation}
which is the limiting value of energy for $T>\Delta $\cite{SS3}.

\section{Discussion}

The main result of this paper is that the parametric dependence of eqs. (\ref
{E_vV_disk-final}), (\ref{E_vV_strip-final})\ and (\ref{E_greater_disk}), (%
\ref{E_greater_strip})\ is the same (with closely matching numerical
coefficients). Consequently, the orbital magnetic response of a
level-quantized metal particle can be satisfactorily explained in terms of
the two-level van Vleck response. This conclusion also holds for other level
distributions characteristic of disordered (chaotic) systems, such as the
Gaussian Unitary Ensemble (only the numerical coefficient will be different).

Whereas the two-level picture might be also valid for the Poisson
distribution, which is the case for classically integrable systems, the mean
response needs to be examined more carefully because level ''bunching,''
characteristic of this distribution, implies that for any magnitude of the
field there will be a significant number of particles for which the
perturbation theory no longer applies.

Another question which we hope to address in a future work is the
sample-specific (fluctuation) effects that are expected to be quite large in
the limit considered here. We point out that, in addition to the fluctuation
of the level spacing and the magnetic moment, the details of the
cancellation between the van Vleck paramagnetism and Landau diamagnetism
should differ from particle to particle leading to variations in the number
of levels contributing to the total response.

\section{Acknowledgment}

I\ wish to thank Bernard Goodman for helpful discussions. This work was not
supported by any funding agency.

\end{document}